\documentclass[]{pasj02} 

\usepackage[switch,mathlines]{lineno} 
\usepackage{siunitx}
\usepackage[whole]{bxcjkjatype} 

\usepackage{comment}

\jyear{2025}
\Received{}
\Accepted{}

\begin{document} 

\title{Simultaneous Formation of the Andromeda Giant Southern Stream and the Substructures in the Andromeda Halo}

\author{
 Misa \textsc{Yamaguchi}\altaffilmark{1}\altemailmark, \orcid{0009-0004-4170-9873} \email{misamisa@ccs.tsukuba.ac.jp} 
 Masao \textsc{Mori}\altaffilmark{2}, \orcid{0000-0001-9033-7085}
 Takanobu \textsc{Kirihara}\altaffilmark{3}, \orcid{0000-0001-6503-8315}
 Yohei \textsc{Miki}\altaffilmark{4}, \orcid{0000-0001-9780-0220}
 Itsuki \textsc{Ogami}\altaffilmark{5, 6}, \orcid{0000-0001-8239-4549}
 Masashi \textsc{Chiba}\altaffilmark{7}, \orcid{0000-0002-9053-860X}
 Yutaka \textsc{Komiyama}\altaffilmark{8}\orcid{0000-0002-3852-6329}
and
Mikito \textsc{Tanaka}\altaffilmark{8}\orcid{0009-0002-7874-5339}
}
\altaffiltext{1}{Graduate School of Science and Technology, University of Tsukuba, 1-1-1 Tennodai, Tsukuba, Ibaraki, 305-8577, Japan}
\altaffiltext{2}{Center for Computational Sciences , University of Tsukuba, 1-1-1 Tennodai, Tsukuba, Ibaraki, 305-8577, Japan}
\altaffiltext{3}{Kitami Institute of Technology, 165, Koen-cho, Kitami, Hokkaido, 090-8507, Japan}
\altaffiltext{4}{Information Technology Center, The University of Tokyo, 6-2-3 Kashiwanoha, Kashiwa, Chiba 277-0882, Japan}
\altaffiltext{5}{The Graduate University for Advanced Studies (SOKENDAI), 2-21-1 Osawa, Mitaka, Tokyo 181-8588, Japan}
\altaffiltext{6}{National Astronomical Observatory of Japan, 2-21-1 Osawa, Mitaka, Tokyo 181-8588, Japan}
\altaffiltext{7}{Astronomical Institute, Tohoku University, Aoba-ku, Sendai, Miyagi 980-8578, Japan}
\altaffiltext{8}{Department of Advanced Sciences, Faculty of Science and Engineering, Hosei University, 3-7-2 Kajino-cho, Koganei, Tokyo 184-8584, Japan}

\KeyWords{galaxies: interactions, galaxies: evolution, galaxies: individual: M31 – Galaxies, galaxies: haloes – Galaxies, Local Group}

\maketitle

\begin{abstract} 
We investigate a minor merger event in M31 that simultaneously forms the Andromeda Giant Southern Stream (AGSS), the Eastern Extent (EE), the North-Eastern Shelf (NES), and the Western Shelf (WS), offering a unified model for these substructures. By varying the scale radius and mass of the dark matter halo associated with the progenitor, around the range predicted by the $\Lambda$ cold dark matter (CDM) model, we successfully reproduce the spatial features of these substructures. Across the limited range of parameters considered in this study, our analysis shows that the spatial evolution of NES and WS is independent of the gravitational potential of the dark matter halo associated with the progenitor, while a shallower potential shifts EE further north. The simulations clearly demonstrate that the progenitor with a dark matter halo mass of $9\times10^9 M_\odot$ colliding with M31 850 Myr ago could simultaneously form the observed substructures, including the AGSS, EE, NES, and WS. The simulation results indicate that EE lies several 10 kpc closer to us than the aligned Stream Cp, which is actually a metal-poor component of Stream C, whose farther distance suggests overlapping debris from distinct collision events, while both remain closely aligned in celestial coordinates. Furthermore, we predict the existence of a positive stream along the AGSS, characterized by positive line-of-sight velocities relative to M31, which complements an already observed negative stream exhibiting negative line-of-sight velocities. Finally, we propose that three objects, namely Stream B, a metal-rich component of Stream C known as Stream Cr, and EE, are components of the Andromeda Giant Southern Arc (AGSA) connected to the AGSS. Although the existence of the positive stream and a complete picture of AGSA have yet to be confirmed observationally, we anticipate that future spectroscopic observations and further advances in theoretical studies will verify their existence.
\end{abstract}

\section{INTRODUCTION}
\label{sec:introduction}
In the $\Lambda$ Cold Dark Matter (CDM) model of galaxy formation, theoretical studies suggest that numerous substructures, such as streams and shells, form within galactic haloes (e.g., \cite{BullockJohnston2005}). Observational studies, including the Pan-Andromeda Archaeological Survey (PAndAS) (e.g., \cite{McConnachie+2009,Lewis+2013, Martin+2013,McConnachie+2018}) and observations made using the Subaru Telescope/Hyper Suprime-Cam \citep{Komiyama+2018,Ogami+2025}, have confirmed the existence of numerous substructures in the halo of M31. These findings support the view that the hierarchical mass assembly process predicted by the $\Lambda$CDM model is still ongoing, as evidenced by both theoretical and observational perspectives. Notable substructures in the halo include the Andromeda Giant Southern Stream (AGSS), the North-Western Stream, Streams A, B, C, and D, the Eastern Extent (EE), the E Cloud, and the SW Cloud. Additionally, the North-Eastern Shelf (NES) and the Western Shelf (WS), located near the disc of M31, are also prominent features \citep{Lewis+2013}.

The AGSS, in particular, has been studied extensively due to its massive and distinctive structure and remains a focal point of research. 
It extends over $\sim 100$ kpc, has a stellar mass of $10^8 \mathrm{M_\odot}$, and exhibits a metallicity of [Fe/H] $\sim -0.5$, originating from a past galaxy collision.
Two distinct models have been proposed for the formation of the AGSS.
Firstly, \citet{Hammer+2010} and \citet{Hammer+2018} propose that M31 underwent a 4:1 major merger. The first passage occurred 7-10 Gyr ago at a pericentre distance of $\qty{32}{kpc}$, with the nuclei coalescing 1.8-3 Gyr ago. In addition, they argue that the AGSS could form under mass ratios ranging from 2:1 to 300:1. 
On the other hand, \citet{Fardal+2007} and \citet{MoriRich2008} proposed a unified model in which a single dwarf galaxy, with a mass below approximately $\qty{5e9}{M_\odot}$, collided to form the AGSS, NES, and WS simultaneously \citep{Fardal+2008,Fardal+2013,Kirihara+2014,Sadoun+2014,Miki+2016,Milosevic+2022,Milosevic+2024}. In the latter scenario, the first pericentric passage occurred within 1 Gyr, shaping the subsequent evolution of the system. This framework also provides insights into the orbital path of the progenitor galaxy given by \citet{Miki+2014} and its morphology given by \citet{Kirihara+2017}.

The observational properties and formation mechanisms of Stream A, Stream B, the E Cloud, and the SW Cloud remain poorly understood. In contrast, so far studies have provided more insight into Streams C and D with \citet{Ibata+2007} initially identifying Streams C and D as thin, metal-poor stream features.
Subsequently, \citet{Chapman2008} identified Stream Cr as a metal-rich component and Stream Cp as a metal-poor component within Stream C, suggesting that Stream C may consist of two distinct components.
\citet{Ibata+2014} also reported a broader, metal-rich structure underlying Stream C. 
From a theoretical perspective, Kaneda et al. (2025) recently proposed that Streams C and D formed through a process independent of the AGSS, using an analytical model and $N$-body simulations.

\citet{Preston+2021} were the first to designate this feature as EE. They revealed that EE is a long, filamentary stellar stream approximately $4^\circ$ (over \qty{100}{kpc}) in length, located \qtyrange[range-phrase = --, range-units = single]{70}{90}{kpc} southeast of the centre of M31. It extends perpendicularly to the minor axis of M31 and overlaps with Stream C. Its radial velocity in the heliocentric frame is $\qty{-368}{\kilo\meter\per\second} \lesssim \ V_\mathrm{los} \lesssim \qty{-331}{\kilo\meter\per\second}$, with a slight velocity gradient of $-0.51 \pm 0.21~\unit{\kilo\meter.\second^{-1}.\kilo pc^{-1}}$. Its photometric metallicity ranges from $-1.0 \lesssim [\mathrm{Fe/H}]_{\mathrm{phot}} \lesssim -0.7$, with an average metallicity of $\langle [\mathrm{Fe/H}]_{\mathrm{phot}} \rangle \sim -0.9$. This is slightly more metal-poor than the AGSS, but the properties of EE and the AGSS are consistent, suggesting that EE may comprise stars stripped from the progenitor galaxy of the AGSS. The stellar population appears to be seamlessly connected to the AGSS. Whether EE is stream-like in nature or represents a unique structure remains an open question.

Despite its vast structure, no theoretical model has yet been proposed to explain the formation of EE, which extends southward from the halo of M31. Previous studies suggested that a substructure might form south of M31 during the formation of the AGSS, as proposed by \citet{Fardal+2008}. However, they expected this substructure to correspond to Streams C and D. Similarly, \citet{Kirihara+2017} suggested the possibility of a comparable structure forming in their simulations, but the observational existence of EE had not been confirmed at the time, and its relevance was not addressed. Thus far, no theoretical model explicitly reproduces the formation of EE.

This situation motivates us to take the next step by building on existing models for the formation of AGSS, NES, and WS within the minor merger framework and developing a new unified model that also accounts for the simultaneous formation of EE.
To achieve this, we propose a framework that reproduces the observed characteristics of the AGSS, the positions of NES and WS, and the spatial distribution of EE. Using $N$-body simulations, we systematically vary the size and mass of the dark matter halo in the progenitor galaxy under the gravitational potential of M31, based on the standard $\Lambda$CDM model. By examining the dynamical evolution of the remnant in detail, we construct a model that successfully explains the simultaneous formation of these structures through a single minor merger event. Furthermore, the model predicts distinct kinematic signatures in phase space, which future spectroscopic observations of the AGSS and EE regions could confirm, providing further validation of this formation scenario.

This paper is organised as follows. Section~\ref{sec:model} describes the properties of the host galaxy, the parameters of the accreting dwarf galaxy, and outlines the $N$-body simulation method. Section~\ref{sec:results} presents the simulation results. Finally, Section~\ref{sec:summary} summarises our conclusions and discusses the prospects for future observability.

\section{SIMULATION MODELS}
\label{sec:model}
In the context of numerical simulations of AGSS formation, \citet{MoriRich2008} conducted a comprehensive $N$-body simulation. In their study, all components of M31—including the bulge, disc, and dark matter halo—along with the progenitor satellite galaxy responsible for the AGSS, were modelled as $N$-body particles. Their findings revealed that the collision of the progenitor satellite galaxy followed a nearly radial infall trajectory, with minimal influence from dynamical friction. This result supports the validity of approximating M31 as a fixed gravitational potential, demonstrating that this simplification does not significantly compromise the accuracy of the collision simulations.

Guided by these insights, we model M31 in this study as a fixed gravitational potential consisting of four components: a dark matter halo described by the NFW profile \citep{NavarroFrenkWhite1996}, a Hernquist bulge \citep{Hernquist1990}, an exponential disc, and a stellar halo represented by a Sersic sphere \citep{Sersic1963}. 
The total mass of the dark matter halo is assumed $\qty{8.11e+11}{M_\odot}$, a scale radius and a virial radius are \qty{7.63}{kpc} and \qty{250}{kpc}, respectively.
The bulge is characterised by a scale radius of \qty{0.610}{kpc}, a scale density of $\qty{2.76e+10}{M_{\odot}.kpc^{-3}}$, and a total mass of $\qty{3.24e+10}{M_{\odot}}$. 
The disc is described by a scale height of \qty{0.600}{kpc}, a scale radius of \qty{5.40}{kpc}, and a total mass of $\qty{3.66e+10}{M_{\odot}}$. 
The inclination and position angles of the M31 disc, as observed on the sky plane from Earth, are set to 77\degree and 37\degree, respectively, as reported by \citet{Geehan+2006}.
The mass, scale radius, and S\'ersic index of the stellar halo are, respectively, $\qty{8.00e+9}{M_\odot}$ \citep{Ibata+2014}, \qty{9.00}{kpc}, and $2.2$ \citep{Gilbert+2012}. 
All of these parameters are adopted from previous studies \citep{Geehan+2006,Fardal+2007,Gilbert+2012,Ibata+2014}.

This study aims to constrain the mass and size of the dark matter halo in the progenitor dwarf galaxy by varying these parameters and developing a unified model for the simultaneous formation of EE and the AGSS. To achieve this, the progenitor galaxy is modelled with three primary components: a dark matter halo characterised by an NFW profile \citep{NavarroFrenkWhite1996}, a King bulge \citep{King1962}, and an exponential disc.
For the dark matter halo of the progenitor, \citet{Kirihara+2017} employed a lowered Evans profile \citep{KuijkenDubinski1994}, which is replaced in this study with the now-standard NFW profile. The parameters of the dark matter halo are determined based on the $c$--$M$ relationship proposed by \citet{Kaneda+2024}. In the model by \citet{Kirihara+2017}, this corresponds to a scale radius of $r_\mathrm{s} = \qty{1.81}{kpc}$, a mass of $\qty{2.84e+9}{M_{\odot}}$, and $c = 18.5$. Using this as the fiducial model, a parameter survey is conducted over the range $c = \numrange{8.50}{68.0}$, with scale radii and total masses ranging from $\qtyrange{0.672}{5.38}{kpc}$ and $\qtyrange{1.40e+9}{5.00e+10}{M_{\odot}}$, respectively.
The virial radius is represented as the density cut-off of the particle distribution.
The bulge is defined with a core radius of \qty{0.227}{kpc}, a dimensionless King parameter of \num{2.05}, and a total mass of $\qty{3.75e+8}{M_{\odot}}$. Disc parameters are based on the reference model proposed by \citet{Kirihara+2017}, with modifications introduced in this study to explore the impact of disc on the formation of substructures. 

In the reference model proposed by \citet{Kirihara+2017}, the disc of the progenitor has a scale height of \qty{0.52}{kpc}, a scale radius of \qty{1.11}{kpc}, and a total mass of $\qty{7.36e+8}{M_{\odot}}$. To improve the model's ability to reproduce the observed characteristics of EE and the AGSS, this study proposes an alternative configuration where the scale height of the disc is reduced to \qty{0.40}{kpc}, while all other parameters remain unchanged. This adjustment allows for an examination of the influence of a thinner disc on the formation of these substructures.

For the initial position and velocity vector of the progenitor, \citet{Fardal+2007} proposed a model that simultaneously forms the AGSS, NES, and WS. Subsequently, \citet{Miki+2014} conducted a comprehensive parameter survey covering the entire northern hemisphere opposite to the AGSS, utilising $5,699,760$ models of the infalling orbit of the progenitor satellite. Their study revealed that the parameters satisfying the conditions for the simultaneous formation of the AGSS, NES, and WS are constrained to a relatively narrow range. Notably, the orbit proposed by \citet{Fardal+2007} lies within this parameter region.

Building on these studies, the coordinate system used in this study aligns the $X$-axis with the minor axis of M31 (northwest), the $Y$-axis with the major axis of M31 (northeast), and the $Z$-axis with the rotation axis of M31 (nearly opposite to the line-of-sight direction).
In this coordinate system, the initial position of the progenitor is $(X, Y, Z) = (\qtylist[list-final-separator = {,}, round-mode=figures, round-precision=3]{22.49669999; -5.44350222; -35.25260661}{kpc})$, and its initial velocity is $(v_X, v_Y, v_Z) = (\qtylist[list-final-separator = {,}, round-mode=figures, round-precision=3]{-28.787987; 19.66586414; 64.68141613}{\kilo\meter\per\second})$. The inclination of the progenitor relative to the plane of the M31 disc is described by the rotation matrix given below:
\begin{equation}
\begin{pmatrix}\label{matrix:prog-rotation}
\num[round-mode=figures, round-precision=3]{0.99102576} & \num[round-mode=figures, round-precision=3]{0.03349337} & \num[round-mode=figures, round-precision=3]{-0.12940687} \\
\num[round-mode=figures, round-precision=3]{0.03349337} & \num[round-mode=figures, round-precision=3]{0.87499714} & \num[round-mode=figures, round-precision=3]{0.48296812} \\
\num[round-mode=figures, round-precision=3]{0.12940687} & \num[round-mode=figures, round-precision=3]{-0.48296812} & \num[round-mode=figures, round-precision=3]{0.8660229} \\
\end{pmatrix},
\end{equation}
as derived by \citet{Kirihara+2017}.

Here, the disc of the progenitor is assumed to rotate counter-clockwise in celestial coordinates, consistent with the model described by \citet{Kirihara+2017}. This rotational behaviour plays a crucial role in shaping the interaction dynamics between the progenitor galaxy and M31, driving the formation of the prominent substructures observed in contemporary data. By incorporating these structural parameters and the dynamics of rotation, this study provides a comprehensive framework for understanding the processes underlying the formation of the AGSS, EE, NES and WS.

The total number of particles distributed by an initial-condition generator 
MAGI (MAny-component Galaxy Initializer) \citep{MikiUmemura2018MAGI} 
is $2^{24} = 16,777,216$, with all particles having equal mass.
To simulate the time evolution of the system, the $N$-body simulation code GOTHIC (Gravitational Oct-Tree code accelerated by HIerarchical time step Controlling) \citep{MikiUmemura2017GOTHIC}
was employed. GOTHIC is a gravitational octree code specifically optimised for graphics processing units (GPU). It utilises the tree method with the accuracy controlling parameter $\varDelta_\mathrm{acc} = 2^{-9}$ to calculate self-gravity on GPU, significantly enhancing computational efficiency. The gravitational softening parameter was set to \qty{15.6}{pc}, and the block time step with the second-order Runge-Kutta method was applied to integrate the equations of motion.

\section{SIMULATION RESULTS}
\label{sec:results}

\subsection{Formation and Evolution of the AGSS and EE}\label{text:time-evol}

\begin{figure}[htbp]
        \begin{center}
	    \includegraphics[width=0.5\textwidth]{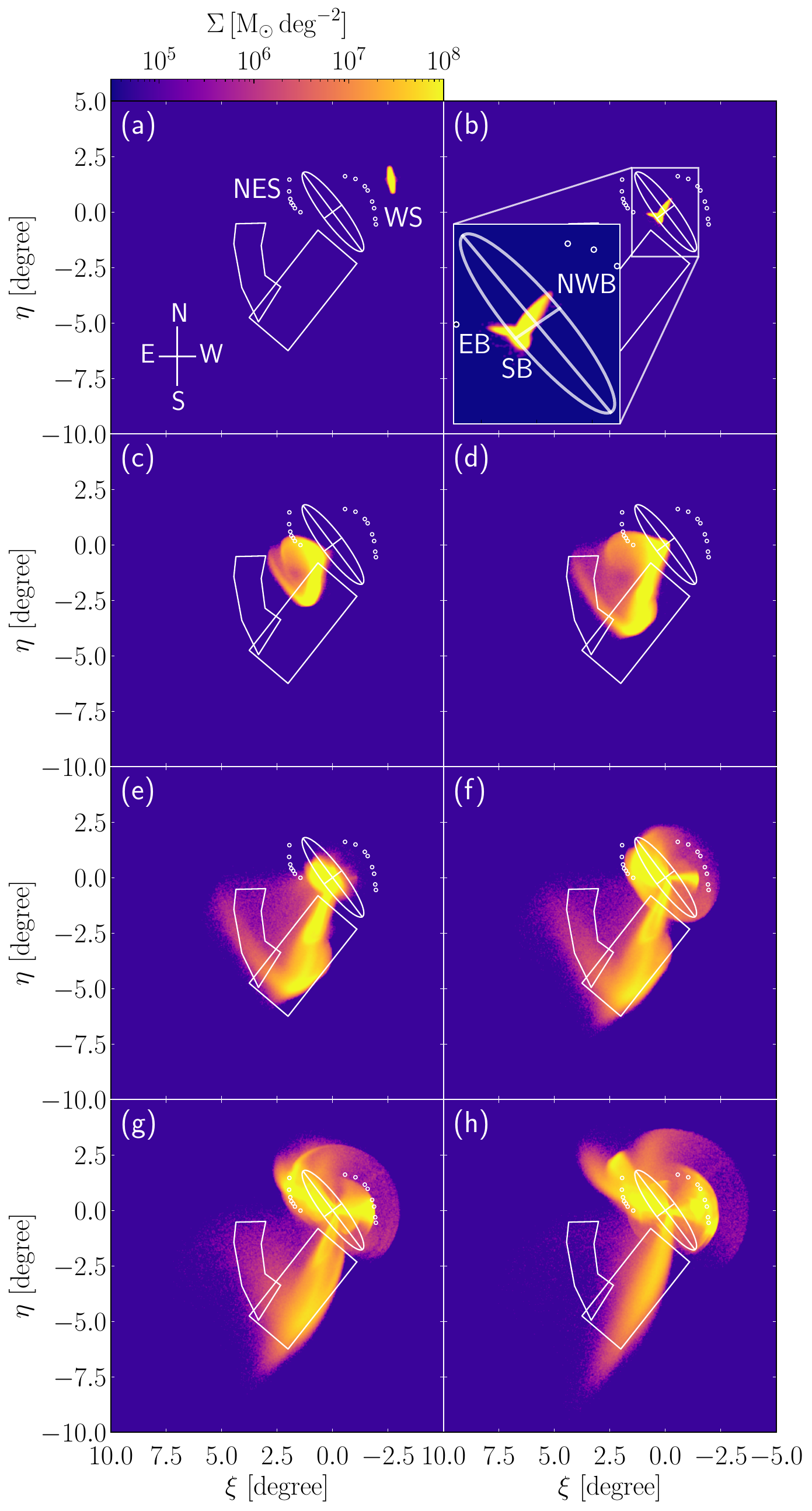}
        \end{center}
        \caption{Time evolution of the stellar surface density distribution in celestial coordinates. The bulge and disc stars of the progenitor are shown undergoing tidal disruption following a central collision with the host galaxy M31. Each panel corresponds to the elapsed time from the start of the simulation: (a) \qty{0.00}{Gyr}, (b) \qty{0.20}{Gyr}, (c) \qty{0.30}{Gyr}, (d) \qty{0.40}{Gyr}, (e) \qty{0.55}{Gyr}, (f) \qty{0.70}{Gyr}, (g) \qty{0.85}{Gyr} (corresponding to the present day), and (h) \qty{1.05}{Gyr}. 
        The white ellipse traces the disc of M31, while the arc-like area enclosed by the white line southeast of M31's centre is defined based on the star count map in \citet{Preston+2021} for comparison with EE. The rectangular area corresponds to the observational fields of the AGSS, as defined by \citet{Conn+2016}. White circles in the eastern and western regions of M31's centre represent the observed locations of the North-Eastern Shelf (NES) and the Western Shelf (WS), respectively \citep{Fardal+2007}. The cardinal directions are displayed in the bottom left corner of panel (a). The Eastern Branch (EB), Southern Branch (SB), North-Western Branch (NWB) are shown in panel (b).
        {Alt text: An eight-panel diagram illustrating the time evolution of the stellar surface mass density distribution in celestial coordinates, depicting tidal disruption during its interaction with M31.}
        }
        \label{fig:time-evol_2-2}
\end{figure}

Figure~\ref{fig:time-evol_2-2} illustrates the time evolution of the tidal disruption of the progenitor in the represented model during the minor merger event, depicted by the surface mass density distribution of the stellar component of the progenitor. In this model, the disc of the progenitor has a dark matter halo with a scale radius of $r_{\mathrm{s}} = \qty{2.69}{kpc}$ and a mass of $M_{\mathrm{DMH}} = \qty{9.0e+9}{M_\odot}$. 
The white circles in Figure~\ref{fig:time-evol_2-2}, arced east and west of M31's centre, indicate the observed NES and WS locations, respectively. The arc-shaped region outlined by a white line southeast of M31's centre marks the observed EE, while the rectangular region to its west represents the AGSS observation area from \citet{Conn+2016}.

Figure~\ref{fig:time-evol_2-2}a shows the start of the simulation, where the progenitor was positioned to the northwest of M31. The progenitor then moved towards the centre of M31, where it was subjected to strong tidal forces exerted by M31's gravitational pull. These forces stretched the progenitor along the north-south direction, significantly altering its structure and redistributing its stellar and dark matter components. 
As illustrated in Figure~\ref{fig:time-evol_2-2}b, after approximately \qty{0.2}{Gyr} from the start of the simulation, the disc of the progenitor completed one full rotation, corresponding to one dynamical time, and reached its first pericentric passage. This event marked the initial collision between the progenitor and M31, leading to significant gravitational interactions and tidal disruption. 

As clearly shown in Figure~\ref{fig:time-evol_2-2}b, the stream separates into a tripod-like structure consisting of three distinct branches: the Eastern Branch (to the left-hand side of the M31 centre), the Southern Branch, and the North-Western Branch. In other words, the stream divides into a leading part that splits into the Eastern Branch and the Southern Branch, and a remaining part directed towards the North-Western Branch. During this collision, the outermost components of the progenitor were the first to pass to the west of the centre of M31. These components were gravitationally deflected eastwards and formed the Eastern Branch. Subsequently, the central core of the progenitor passed to the east of the centre of M31, was deflected westwards, and developed into the Southern Branch. Finally, the existing North-Western Branch is expected to pass to the east of the centre of M31. 
As discussed later, the Eastern Branch is expected to move rapidly northward, and the South Branch and North-Western Branch develop into the AGSS and EE.

Figure~\ref{fig:time-evol_2-2}c, at \qty{0.3}{Gyr}, shows a faint arc-like structure in the south-eastern part of the M31 halo. This structure arises from gravitational scattering caused by a portion of the North-Western Branch entering the region east of the centre of M31. These particles orbit around the centre of M31 and then start moving in a south-eastern direction.
On the other hand, the anti-clockwise rotation of the disc of the progenitor in celestial coordinates plays a key role in enhancing the particle wraparound effect and increasing the number of particles entering the region east of the centre of M31. By \qty{0.4}{Gyr}, the arc-like structure becomes more distinct in Figure~\ref{fig:time-evol_2-2}d, gradually expanding and extending southeastwards. Meanwhile, the bulk of the North-Western Branch passes to the east of the centre of M31 and begins to form the root of the AGSS.

In Figure~\ref{fig:time-evol_2-2}e, corresponding to \qty{0.55}{Gyr}, the Eastern Branch scatters northwards and begins to form a shell structure that fans out to the north of the M31 centre. Shortly afterwards, the core of the progenitor undergoes its second pericentric passage. Furthermore, the expansion rate of the arc-like structure towards the south-east gradually slows down, the constituent particles nearly reach their respective apocentres, and an EE-like structure emerges.
At \qty{0.7}{Gyr}, Figure~\ref{fig:time-evol_2-2}f illustrates that particles in the Eastern Branch have already passed their northeastern apocentres. Subsequently, these particles traverse near the centre of M31 before continuing northwest along their orbits, with the most advanced particles already surpassing the observed location of WS. Meanwhile, the majority of particles in the North-Western Branch follow a path through the AGSS, contributing to the formation of NES, while the leading group progresses to form WS.

Figure~\ref{fig:time-evol_2-2}g, at \qty{0.85}{Gyr}, corresponds to the present day. It is evident that the NES and WS structures are fully reproduced on the northern side. Conversely, in the south, the AGSS structure is well developed and aligns perfectly with observations. Furthermore, EE, extending into an arc-like structure slightly north of the apogee, is also nearly fully reproduced. This indicates that all four observed characteristic structures could be reproduced simultaneously in a single minor collision event.
In \qty{0.2}{Gyr} from the present day, at \qty{1.05}{Gyr} from the start of the simulation, a third pericentric transit occurs, during which the stars forming the AGSS and EE return to the centre of M31, causing the EE-like structures to disappear while the AGSS remains intact (Figure~\ref{fig:time-evol_2-2}h). Meanwhile, the northern shelves continue to expand further into the outer regions.

\begin{figure}[t]
        \begin{center}
	    \includegraphics[width=0.5\textwidth]{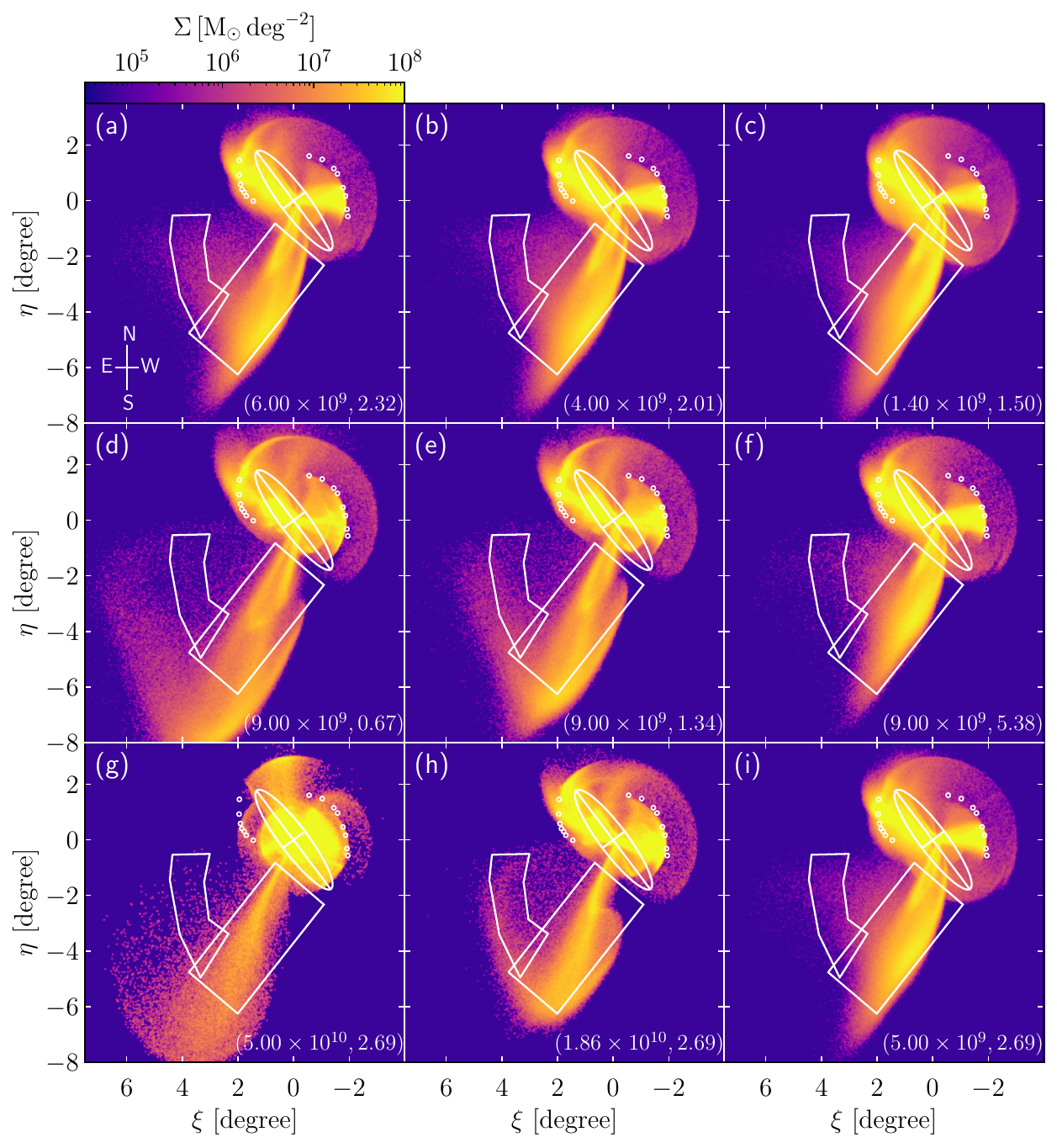}
        \end{center}
        \caption{Distribution of the stellar surface density resulting from varying dark matter halo properties in the progenitor. All snapshots correspond to the present day (0.85 Gyr from the start of the simulation). 
        The white symbols follow the conventions of Figure \ref{fig:time-evol_2-2}.
        Panels (a–c): Results for varying both the scale radius and the mass of the dark matter halo of the progenitor, following the $\Lambda$CDM model. Panels (d–f): Results for varying the scale radius while keeping the mass and virial radius fixed. Panels (g–i): Results for varying the mass while keeping the scale radius and virial radius fixed.
        {Alt text: A nine-panel diagram showing the distribution of the stellar surface mass density. The diagrams illustrate the dynamical impact of varying dark matter halo parameters on the spatial configuration of substructures.}
        }
        \label{fig:parameter-survey}
\end{figure}

\subsection{Impact of the Dark Matter Halo in the Progenitor} \label{text:experiment}

In this section, we present the results of numerical experiments that investigate how the dark matter halo associated with a progenitor galaxy after its collision with M31 affects the spatial evolution of the substructure of the remnant of the progenitor galaxy.

It is widely known that dark matter haloes have a universal mass density distribution function in the $\Lambda$CDM model, and furthermore, it has been theoretically and observationally verified that the mass distribution function itself follows a scaling law with respect to its total mass \citep[and references therein]{BullockBoylan-Kolchin2017}. Assuming an NFW profile for the mass distribution
\begin{equation}
M(r)=M_\mathrm{DMH} \frac{\log(1+r/r_\mathrm{s})
-(r/r_\mathrm{s})/(1+r/r_\mathrm{s})}{\log(1+c)-c/(1+c)},
\end{equation}
where $M_\mathrm{DMH}$ is the total mass of the dark matter halo within a virial radius $r_\mathrm{vir}=r_\mathrm{vir}(M_\mathrm{vir})$, where $M_\mathrm{vir}$ is the total mass of the system. 
\citet{Moline+2023} and \citet{Kaneda+2024} derived an expression linking the concentration parameter $c$ to $M_\mathrm{vir}$ by analysing results from ultra-high-resolution cosmological simulations. Here, we adopt the relationship $c = c(M_\mathrm{vir})$ as provided by \citet{Kaneda+2024}.
The scale length $r_\mathrm{s}$ is given by $r_\mathrm{s}=r_\mathrm{s}(M_\mathrm{vir})=r_\mathrm{vir}(M_\mathrm{vir})/c(M_\mathrm{vir})$.

Following these guidelines, this analysis incorporates the scaling laws of the dark matter halo predicted by the $\Lambda$CDM model, along with potential deviations from these laws. Numerical experiments are performed by varying the mass of the dark matter halo from $\qty{5e+9}{M_\odot}$ to $\qty{5e+10}{M_\odot}$ and the scale length from \qty{0.67}{kpc} to \qty{5.38}{kpc}, while keeping the mass and scale length of the stellar components fixed, except for the disc thickness, which adjusts according to the gravitational potential. The results of these experiments are shown in Figure~\ref{fig:parameter-survey}.

Figure~\ref{fig:parameter-survey}a–c presents the results for the $\Lambda$CDM track, where both the scale radius and the mass of the dark matter halo associated with the progenitor are varied while following the $\Lambda$CDM conditions \citep{Kaneda+2024}. The results for several parameter sets are displayed in these panels. At an elapsed time of \qty{0.85}{Gyr} from the start of the simulation, NES and WS are consistently located on the northern side for all parameter sets. However, as the mass decreases, the system evolves more rapidly due to easier tidal disruption caused by the shallower gravitational potential of the progenitor, which weakens its overall binding. Consequently, EE gradually shifts northwards.

Next, we examine the case where the mass and virial radius of the dark matter halo associated with the progenitor are fixed, while the scale radius is varied. The results for several parameter sets are shown in Figure~\ref{fig:parameter-survey}d–f. Similar to the previous results, the spatial distribution of the two northern shelves is reproduced at \qty{0.85}{Gyr}, even when the scale radius of the dark matter halo varies significantly. The AGSS and EE shift further north as the scale radius increases, as shown from Figure~\ref{fig:parameter-survey}d to f, again suggesting that a shallower gravitational potential of the progenitor influences the formation of the AGSS and EE.

Finally, we investigate the case where the scale radius and virial radius of the dark matter halo are fixed, and the mass is varied. The results for several parameter sets are shown in Figure~\ref{fig:parameter-survey}g-i. Interestingly, with the exception of Figure~\ref{fig:parameter-survey}g, which corresponds to the most massive halo in the considered range, the two northern shelves are successfully reproduced at \qty{0.85}{Gyr} in all cases. This outcome likely arises because, during the first pericentric passage, the progenitor's particles become unbound from its gravitational field and move as free particles within the fixed potential field of M31. The EE-like structure on the southern side shifts northwards as the mass of the dark matter halo associated with the progenitor decreases. Conversely, the length of the AGSS shows only a weak correlation with the mass of the dark matter halo. 

In Figure~\ref{fig:parameter-survey}g, no structure corresponding to EE is present, the two northern shelves are not fully extended, and the shape of the AGSS is inconsistent with observations. This suggests that the strong gravitational constraints imposed by the dark matter halo associated with the progenitor significantly limit the motion of the stars, thereby preventing the formation of diverse substructures. Therefore, we conclude that neither EE nor the AGSS, along with NES and WS, can be reproduced with a dark matter halo mass exceeding $\qty{5e+10}{M_\odot}$, indicating that such massive dark matter haloes fail to reproduce the observed features, a finding consistent with the results of \citet{MoriRich2008}.

\subsection{Signatures of Substructures on Phase Space}\label{text:pahse space}
    
\begin{figure}[t]
        \begin{center}
	    \includegraphics[width=0.5\textwidth]{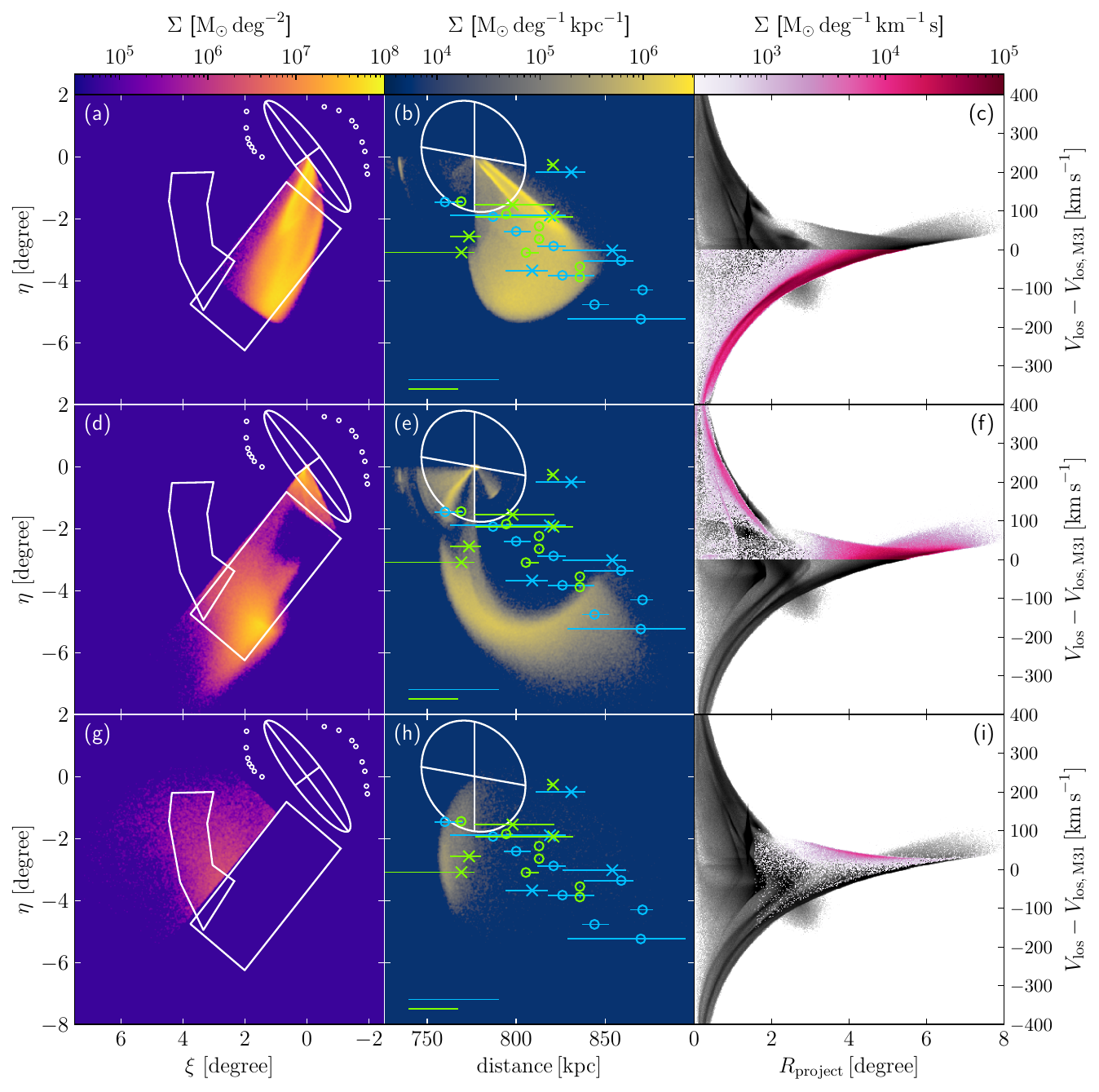}
        \end{center}
        \caption{Spatial and phase space distributions of the resultant stellar components. Panels a–c display the distribution of stellar particles in the AGSS region with their line-of-sight velocities $V_{\mathrm{los}} - V_{\mathrm{los, \,M31}} \leq 0 \,\mathrm{km \,s^{-1}}$ in the heliocentric coordinate system. Panels d–f show those with $V_{\mathrm{los}} - V_{\mathrm{los, \,M31}} > 0 \,\mathrm{km \,s^{-1}}$, while panels g–i show the distribution in the EE region. The white symbols follow the conventions of Figure \ref{fig:time-evol_2-2}.
        The AGSS and EE regions are defined as $\eta < -\,\xi$ and $\eta < 2.2\,\xi$, and $\eta \ge -\,\xi$ and $\eta < \xi - 2.2^{\circ}$, respectively. Panels a, d, and g show the celestial coordinate plane ($\xi$, $\eta$), panels b, e, and h show $\eta$ and $\mathrm{distance}$, and panels c, f, and i show $V_{\mathrm{los}} - V_{\mathrm{los, \,M31}}$ as a function of the projected distance from the centre of M31 ($R_{\mathrm{project}}$). The blue symbols in panels b, e, and h represent the observed distances to the AGSS and Stream C, respectively, as reported by \citet{Conn+2016}, while the green symbols correspond to data from \citet{Ogami+2025}. Open circles (crosses) represent the observed distances along the AGSS (Stream C). The horizontal lines in the bottom-left corner of these panels indicate systematic errors in the observations, with green representing \citep{Ogami+2025} and blue representing \citep{Conn+2016}. Grey-scale maps in panels c, f, and i show the distribution of stellar particles from the progenitor in the entire M31 halo, as presented in Figures~\ref{fig:time-evol_2-2}. 
        {Alt text: A nine-panel scientific diagram visualising the distribution of progenitor disc particles in the AGSS and EE regions of the M31 halo. The diagrams include celestial coordinates, $\eta$ versus distance, and line-of-sight velocity versus projected distance from M31's centre. Colour-coded symbols and grey-scale maps provide additional data overlays, distinguishing observed and simulated features.}
        }
        \label{fig:AGSS_EE}
\end{figure}

In this section, we analyse the relationship between the distributions of particles forming the AGSS and those forming EE in phase space, by selecting particles based on their respective regions. For this purpose, two sampling regions are defined: the AGSS region ($\eta < -\xi$ and $\eta < 2.2\xi$) and the EE region ($\eta \ge -\xi$ and $\eta < \xi - 2.2^{\circ}$).

Figures~\ref{fig:AGSS_EE}a–c show the stellar particles from the progenitor in the AGSS region with line-of-sight velocities $V_\mathrm{los} \leq V_\mathrm{los,\, M31}$, referred to as the negative stream, in the heliocentric coordinate system, where the line-of-sight velocity of M31 $V_\mathrm{los,\, M31}$ is $\qty{-300}{km.s^{-1}}$ \citep{deVaucouleurs1991}. Figures~\ref{fig:AGSS_EE}d–f show the stellar particles in the AGSS region with $V_\mathrm{los} > V_\mathrm{los,\, M31}$, referred to as the positive stream, while Figures~\ref{fig:AGSS_EE}g–i depict the mass distribution of stellar particles in the EE region across different coordinate planes.
Figures~\ref{fig:AGSS_EE}a, d, and g represent the celestial coordinate plane ($\xi$-$\eta$). Figures~\ref{fig:AGSS_EE}b, e, and h show the relationship between $\eta$ and the distance from Earth ($\mathrm{distance}$), and Figures~\ref{fig:AGSS_EE}c, f, and i illustrate the dependence of the line-of-sight velocity $V_\mathrm{los}$ on the projected distance from the centre of M31 ($R_\mathrm{project}$). The background in these panels, represented in grayscale, shows the line-of-sight velocity maps for all stellar particles across the entire M31 halo.

In Fig.~\ref{fig:AGSS_EE}b, along the AGSS, there seems to be a fair level of agreement between the observational data and our simulation.  In contrast, significant variation is already evident in the observations of Stream C in Figs.~\ref{fig:AGSS_EE}b, e and h.  According to the simulation, Stream C and EE overlap in the depth direction within the celestial coordinate system, suggesting the potential contamination between observed two structures. 
We propose that Stream C, especially the metal poor component of Stream Cp, arose from a different galaxy collision event and investigate its possible formation with Stream D; however, the details are presented elsewhere (i.e., Kaneda et al. 2025).

From Figures~\ref{fig:AGSS_EE}a, b, and c, the dense regions of negative stream components visible in Figure~\ref{fig:AGSS_EE}c correspond to the AGSS. Conversely, Figures~\ref{fig:AGSS_EE}d, e, and f reveal the presence of positive stream components in Figure~\ref{fig:AGSS_EE}f. Among these, the components with $R_{\mathrm{project}} < 2^{\circ}$, formed by structures near the centre of M31, correspond to part of the shell near the M31 disc, extending slightly westward from the centre of M31.
On the other hand, the positive stream components with $R_{\mathrm{project}} > 2.5^{\circ}$ in Figure~\ref{fig:AGSS_EE}f, formed by structures further from the centre of M31, are associated with an arc-like structure leading to EE. This arc extends from the west to the south of the AGSS, as described in Section~\ref{text:time-evol}. 
Furthermore, from Figure~\ref{fig:AGSS_EE}i, it is evident that EE forms part of the $R_{\mathrm{project}} > 2.5^{\circ}$ component of the positive stream seen in Figure~\ref{fig:AGSS_EE}f.
The observed line-of-sight velocities of the EE region, reported by \citet{Preston+2021}, range from $\qty{-368}{km.s^{-1}}$ to $\qty{-331}{km.s^{-1}}$ in the heliocentric coordinate system. Comparing these with Figure~\ref{fig:AGSS_EE}i, their results correspond to $\qty{-68}{km.s^{-1}} \lesssim V_\mathrm{los} \lesssim \qty{-31}{km.s^{-1}}$ in the M31-centred coordinate system, indicating that they are not part of the positive stream.

The existence of the positive stream described here has not been previously reported. Considering the time evolution of substructures discussed in Section~\ref{text:time-evol}, one of the main components of the positive stream seen in Figure~\ref{fig:AGSS_EE}i is likely part of EE currently moving away from us and returning towards the centre of M31. Additionally, it is found that the AGSS contains both a negative stream, with negative line-of-sight velocity components moving towards us, and a positive stream, currently moving away from us. 

\section{SUMMARY AND DISCUSSIONS}
\label{sec:summary}

This study demonstrates that the minor merger responsible for the formation of the AGSS can also simultaneously form EE, NES, and WS.
By varying the scale radius and mass of the dark matter halo within the range predicted by the $\Lambda$CDM model, accounting for deviations from it, and adjusting the gravitational potential, the observed EE was successfully reproduced qualitatively. Consequently, we have established a model for the simultaneous formation of four distinct substructures in the halo of M31, namely the AGSS, EE, NES, and WS. In other words, the infalling dwarf galaxies responsible for the collision were formed from dark matter haloes that originated from the average density fluctuations in the $\Lambda$CDM universe.
Additionally, this study revealed that the temporal evolution of the spatial distribution of NES and WS is largely independent of the gravitational potential of the dark matter halo of the progenitor within the given parameter ranges. Moreover, a shallower gravitational potential results in EE forming further to the north. These physical mechanisms will be explored in greater detail in future studies.

\citet{Fardal+2008} demonstrated that a substructure appeared in the south-eastern direction of the M31 halo when the rotation of the disc on the celestial plane was clockwise at the initial condition. In contrast, \citet{Kirihara+2017} successfully reproduced the internal structure of the AGSS, including the asymmetrical surface brightness distribution in the azimuthal direction, by assuming an anti-clockwise disc. Their results also indicated the presence of substructure in the south-eastern region of the M31 halo. However, at that time, the existence of EE itself was not observationally confirmed, and detailed information regarding its spatial structure and heavy element content remained unavailable. In this study, we further refine the counterclockwise disc model to explain the spatial distribution of EE by adjusting the structure of the dark matter halo associated with the progenitor. Future investigations are anticipated to reveal differences in rotational direction within the velocity and metallicity distributions of collision remnants in phase space, providing deeper insights into the dynamics of such interactions.

Figure \ref{fig:AGSS_EE} reveals that the positive stream consists of three distinct components: 
(A) a part of EE moving northward in the southeast toward M31's centre (lower panel), (B) a part of EE extending along the line-of-sight depth in the southern AGSS and connecting to it (middle panel, $R_{\text{proj}} > 5$), and (C) a section of the AGSS overlapping the negative stream and moving southward toward M31's centre (middle panel, $3 < R_{\text{proj}} < 5$). 
These results suggest that component (A) should be observable along the EE as an independent structure from the AGSS. In contrast, component (B) appears linked to the negative stream along the AGSS, both originating near M31's centre.
In \citet{Dey2023}, a clear negative stream was observed, but no evidence of a positive stream was reported. This absence may reflect an observational bias towards metal-rich stars, potentially neglecting the metal-poor component. Future spectroscopic surveys such as by Subaru Prime Focus Spectrograph (PFS) \citep{Takada2014, Tamura2024} targeting the EE and AGSS-related substructures could confirm its existence. Such observations, combined with advancements in high-resolution imaging and spectroscopy, will refine our understanding of the gravitational potential of the progenitor, the formation mechanisms of the AGSS and EE, and the relationships between these substructures in the halo of M31, offering deeper insights into the history of the galaxy assembly.

Numerous substructures have been discovered in the halo of M31, with many complex features reported in the southern halo. However, the origins of these structures, whether independent or interrelated, are often discussed based on spectroscopic observations, such as line-of-sight velocities and metallicities. Due to the extreme difficulty of such observations, typically only a few dozen stars can be studied spectroscopically within a limited field of view. Consequently, the details of each substructure and their origins remain only partially understood. 
For instance, \citet{Preston+2021} reported that the metallicity of EE is consistent with that of Stream Cr and that its line-of-sight velocity coincides with that of Stream B. They further speculated on a potential relationship between Stream B and Stream Cr in the formation of EE. 

Here, we propose a bold hypothesis: if Stream B, Stream Cr, and EE are part of the same celestial structure during the formation epoch of the AGSS, this would imply the existence of a massive stellar aggregation, which we tentatively refer to as the Andromeda Giant Southern Arc (AGSA). This structure, comparable in scale to the AGSS, would be an extremely faint yet enormous feature extending over 100 kpc in length and tens of kpc in width in the southeastern part of the M31 halo. The huge structure observed in the southeastern region of our simulation results, connecting to the southern end of the AGSS, strongly suggests a potential relationship with the AGSA, and we cannot stop reflecting on the profound possibility of their link. Further verification is necessary to substantiate this hypothesis. In the future, achieving greater statistical reliability will require acquiring and analysing spectroscopic data for a larger number of stars across wider observational areas. Additionally, constructing more precise models that can be compared in greater detail with observational data, including metallicity and radial velocity distributions of the AGSA, will be crucial for drawing significant conclusions.

\begin{ack}
We thank Janet Preston and an anonymous reviewer for valuable comments and suggestions, which have helped improve the quality of this manuscript.
This work was supported by JSPS KAKENHI Grant Numbers JP24K07085, JP24K00669, JP23K11123, JP22K14076 and JP20K14517. Numerical computations were performed with computational resources provided by the Multidisciplinary Cooperative Research Program in the Center for Computational Sciences, University of Tsukuba.
\end{ack}

\bibliographystyle{apj}
\bibliography{ref}

\begin{thebibliography}{}
\expandafter\ifx\csname natexlab\endcsname\relax\def\natexlab#1{#1}\fi

\bibitem[{{Bullock} \& {Boylan-Kolchin}(2017)}]{BullockBoylan-Kolchin2017}
{Bullock}, J.~S., \& {Boylan-Kolchin}, M. 2017, \araa, 55, 343

\bibitem[{{Bullock} \& {Johnston}(2005)}]{BullockJohnston2005}
{Bullock}, J.~S., \& {Johnston}, K.~V. 2005, \apj, 635, 931

\bibitem[{{Chapman} {et~al.}(2008){Chapman}, {Ibata}, {Irwin}, {Koch}, {Letarte}, {Martin}, {Collins}, {Lewis}, {McConnachie}, {Pe{\~n}arrubia}, {Rich}, {Trethewey}, {Ferguson}, {Huxor}, \& {Tanvir}}]{Chapman2008}
{Chapman}, S.~C., {Ibata}, R., {Irwin}, M., {et~al.} 2008, \mnras, 390, 1437

\bibitem[{{Conn} {et~al.}(2016){Conn}, {McMonigal}, {Bate}, {Lewis}, {Ibata}, {Martin}, {McConnachie}, {Ferguson}, {Irwin}, {Elahi}, {Venn}, \& {Mackey}}]{Conn+2016}
{Conn}, A.~R., {McMonigal}, B., {Bate}, N.~F., {et~al.} 2016, \mnras, 458, 3282

\bibitem[{{de Vaucouleurs} {et~al.}(1991){de Vaucouleurs}, {de Vaucouleurs}, {Corwin}, {Buta}, {Paturel}, \& {Fouque}}]{deVaucouleurs1991}
{de Vaucouleurs}, G., {de Vaucouleurs}, A., {Corwin}, Jr., H.~G., {et~al.} 1991, {Third Reference Catalogue of Bright Galaxies}

\bibitem[{{Dey} {et~al.}(2023){Dey}, {Najita}, {Koposov}, {Josephy-Zack}, {Maxemin}, {Bell}, {Poppett}, {Patel}, {Beraldo e Silva}, {Raichoor}, {Schlegel}, {Lang}, {Meisner}, {Myers}, {Aguilar}, {Ahlen}, {Allende Prieto}, {Brooks}, {Cooper}, {Dawson}, {de la Macorra}, {Doel}, {Font-Ribera}, {Garc{\'\i}a-Bellido}, {Gontcho A Gontcho}, {Guy}, {Honscheid}, {Kehoe}, {Kisner}, {Kremin}, {Landriau}, {Le Guillou}, {Levi}, {Li}, {Martini}, {Miquel}, {Moustakas}, {Nie}, {Palanque-Delabrouille}, {Prada}, {Schlafly}, {Sharples}, {Tarl{\'e}}, {Ting}, {Tyas}, {Valluri}, {Wechsler}, \& {Zou}}]{Dey2023}
{Dey}, A., {Najita}, J.~R., {Koposov}, S.~E., {et~al.} 2023, \apj, 944, 1

\bibitem[{{Fardal} {et~al.}(2008){Fardal}, {Babul}, {Guhathakurta}, {Gilbert}, \& {Dodge}}]{Fardal+2008}
{Fardal}, M.~A., {Babul}, A., {Guhathakurta}, P., {Gilbert}, K.~M., \& {Dodge}, C. 2008, \apjl, 682, L33

\bibitem[{{Fardal} {et~al.}(2007){Fardal}, {Guhathakurta}, {Babul}, \& {McConnachie}}]{Fardal+2007}
{Fardal}, M.~A., {Guhathakurta}, P., {Babul}, A., \& {McConnachie}, A.~W. 2007, \mnras, 380, 15

\bibitem[{{Fardal} {et~al.}(2013){Fardal}, {Weinberg}, {Babul}, {Irwin}, {Guhathakurta}, {Gilbert}, {Ferguson}, {Ibata}, {Lewis}, {Tanvir}, \& {Huxor}}]{Fardal+2013}
{Fardal}, M.~A., {Weinberg}, M.~D., {Babul}, A., {et~al.} 2013, \mnras, 434, 2779

\bibitem[{{Geehan} {et~al.}(2006){Geehan}, {Fardal}, {Babul}, \& {Guhathakurta}}]{Geehan+2006}
{Geehan}, J.~J., {Fardal}, M.~A., {Babul}, A., \& {Guhathakurta}, P. 2006, \mnras, 366, 996

\bibitem[{{Gilbert} {et~al.}(2012){Gilbert}, {Guhathakurta}, {Beaton}, {Bullock}, {Geha}, {Kalirai}, {Kirby}, {Majewski}, {Ostheimer}, {Patterson}, {Tollerud}, {Tanaka}, \& {Chiba}}]{Gilbert+2012}
{Gilbert}, K.~M., {Guhathakurta}, P., {Beaton}, R.~L., {et~al.} 2012, \apj, 760, 76

\bibitem[{{Hammer} {et~al.}(2018){Hammer}, {Yang}, {Wang}, {Ibata}, {Flores}, \& {Puech}}]{Hammer+2018}
{Hammer}, F., {Yang}, Y.~B., {Wang}, J.~L., {et~al.} 2018, \mnras, 475, 2754

\bibitem[{{Hammer} {et~al.}(2010){Hammer}, {Yang}, {Wang}, {Puech}, {Flores}, \& {Fouquet}}]{Hammer+2010}
---. 2010, \apj, 725, 542

\bibitem[{{Hernquist}(1990)}]{Hernquist1990}
{Hernquist}, L. 1990, \apj, 356, 359

\bibitem[{{Ibata} {et~al.}(2007){Ibata}, {Martin}, {Irwin}, {Chapman}, {Ferguson}, {Lewis}, \& {McConnachie}}]{Ibata+2007}
{Ibata}, R., {Martin}, N.~F., {Irwin}, M., {et~al.} 2007, \apj, 671, 1591

\bibitem[{{Ibata} {et~al.}(2014){Ibata}, {Lewis}, {McConnachie}, {Martin}, {Irwin}, {Ferguson}, {Babul}, {Bernard}, {Chapman}, {Collins}, {Fardal}, {Mackey}, {Navarro}, {Pe{\~n}arrubia}, {Rich}, {Tanvir}, \& {Widrow}}]{Ibata+2014}
{Ibata}, R.~A., {Lewis}, G.~F., {McConnachie}, A.~W., {et~al.} 2014, \apj, 780, 128

\bibitem[{{Kaneda} {et~al.}(2024){Kaneda}, {Mori}, \& {Otaki}}]{Kaneda+2024}
{Kaneda}, Y., {Mori}, M., \& {Otaki}, K. 2024, \pasj, 76, 1026

\bibitem[{{King}(1962)}]{King1962}
{King}, I. 1962, \aj, 67, 471

\bibitem[{{Kirihara} {et~al.}(2014){Kirihara}, {Miki}, \& {Mori}}]{Kirihara+2014}
{Kirihara}, T., {Miki}, Y., \& {Mori}, M. 2014, \pasj, 66, L10

\bibitem[{{Kirihara} {et~al.}(2017){Kirihara}, {Miki}, {Mori}, {Kawaguchi}, \& {Rich}}]{Kirihara+2017}
{Kirihara}, T., {Miki}, Y., {Mori}, M., {Kawaguchi}, T., \& {Rich}, R.~M. 2017, \mnras, 464, 3509

\bibitem[{{Komiyama} {et~al.}(2018){Komiyama}, {Chiba}, {Tanaka}, {Tanaka}, {Kirihara}, {Miki}, {Mori}, {Lupton}, {Guhathakurta}, {Kalirai}, {Gilbert}, {Kirby}, {Lee}, {Jang}, {Sharma}, \& {Hayashi}}]{Komiyama+2018}
{Komiyama}, Y., {Chiba}, M., {Tanaka}, M., {et~al.} 2018, \apj, 853, 29

\bibitem[{{Kuijken} \& {Dubinski}(1994)}]{KuijkenDubinski1994}
{Kuijken}, K., \& {Dubinski}, J. 1994, \mnras, 269, 13

\bibitem[{{Lewis} {et~al.}(2013){Lewis}, {Braun}, {McConnachie}, {Irwin}, {Ibata}, {Chapman}, {Ferguson}, {Martin}, {Fardal}, {Dubinski}, {Widrow}, {Mackey}, {Babul}, {Tanvir}, \& {Rich}}]{Lewis+2013}
{Lewis}, G.~F., {Braun}, R., {McConnachie}, A.~W., {et~al.} 2013, \apj, 763, 4

\bibitem[{{Martin} {et~al.}(2013){Martin}, {Ibata}, {McConnachie}, {Mackey}, {Ferguson}, {Irwin}, {Lewis}, \& {Fardal}}]{Martin+2013}
{Martin}, N.~F., {Ibata}, R.~A., {McConnachie}, A.~W., {et~al.} 2013, \apj, 776, 80

\bibitem[{{McConnachie} {et~al.}(2009){McConnachie}, {Irwin}, {Ibata}, {Dubinski}, {Widrow}, {Martin}, {C{\^o}t{\'e}}, {Dotter}, {Navarro}, {Ferguson}, {Puzia}, {Lewis}, {Babul}, {Barmby}, {Bienaym{\'e}}, {Chapman}, {Cockcroft}, {Collins}, {Fardal}, {Harris}, {Huxor}, {Mackey}, {Pe{\~n}arrubia}, {Rich}, {Richer}, {Siebert}, {Tanvir}, {Valls-Gabaud}, \& {Venn}}]{McConnachie+2009}
{McConnachie}, A.~W., {Irwin}, M.~J., {Ibata}, R.~A., {et~al.} 2009, \nat, 461, 66

\bibitem[{{McConnachie} {et~al.}(2018){McConnachie}, {Ibata}, {Martin}, {Ferguson}, {Collins}, {Gwyn}, {Irwin}, {Lewis}, {Mackey}, {Davidge}, {Arias}, {Conn}, {C{\^o}t{\'e}}, {Crnojevic}, {Huxor}, {Penarrubia}, {Spengler}, {Tanvir}, {Valls-Gabaud}, {Babul}, {Barmby}, {Bate}, {Bernard}, {Chapman}, {Dotter}, {Harris}, {McMonigal}, {Navarro}, {Puzia}, {Rich}, {Thomas}, \& {Widrow}}]{McConnachie+2018}
{McConnachie}, A.~W., {Ibata}, R., {Martin}, N., {et~al.} 2018, \apj, 868, 55

\bibitem[{{Miki} {et~al.}(2014){Miki}, {Mori}, {Kawaguchi}, \& {Saito}}]{Miki+2014}
{Miki}, Y., {Mori}, M., {Kawaguchi}, T., \& {Saito}, Y. 2014, \apj, 783, 87

\bibitem[{{Miki} {et~al.}(2016){Miki}, {Mori}, \& {Rich}}]{Miki+2016}
{Miki}, Y., {Mori}, M., \& {Rich}, R.~M. 2016, \apj, 827, 82

\bibitem[{{Miki} \& {Umemura}(2017)}]{MikiUmemura2017GOTHIC}
{Miki}, Y., \& {Umemura}, M. 2017, New Astronomy, 52, 65

\bibitem[{Miki \& Umemura(2018)}]{MikiUmemura2018MAGI}
Miki, Y., \& Umemura, M. 2018, MNRAS, 475, 2269^^e2^^80^^932281

\bibitem[{{Milo{\v{s}}evi{\'c}} {et~al.}(2022){Milo{\v{s}}evi{\'c}}, {Mi{\'c}i{\'c}}, \& {Lewis}}]{Milosevic+2022}
{Milo{\v{s}}evi{\'c}}, S., {Mi{\'c}i{\'c}}, M., \& {Lewis}, G.~F. 2022, \mnras, 511, 2868

\bibitem[{{Milo{\v{s}}evi{\'c}} {et~al.}(2024){Milo{\v{s}}evi{\'c}}, {Mi{\'c}i{\'c}}, \& {Lewis}}]{Milosevic+2024}
---. 2024, \mnras, 527, 4797

\bibitem[{Molin^^c3^^a9 {et~al.}(2023)Molin^^c3^^a9, S^^c3^^a1nchez-Conde, Aguirre-Santaella, Ishiyama, Prada, Cora, Croton, Jullo, Metcalf, Oogi, \& Ruedas}]{Moline+2023}
Molin^^c3^^a9, {\rm{\'{A}}}., S^^c3^^a1nchez-Conde, M.~A., Aguirre-Santaella, A., {et~al.} 2023, \mnras, 518, 157

\bibitem[{{Mori} \& {Rich}(2008)}]{MoriRich2008}
{Mori}, M., \& {Rich}, R.~M. 2008, \apjl, 674, L77

\bibitem[{{Navarro} {et~al.}(1996){Navarro}, {Frenk}, \& {White}}]{NavarroFrenkWhite1996}
{Navarro}, J.~F., {Frenk}, C.~S., \& {White}, S. D.~M. 1996, \apj, 462, 563

\bibitem[{{Ogami} {et~al.}(2025){Ogami}, {Tanaka}, {Komiyama}, {Chiba}, {Guhathakurta}, {Kirby}, {Wyse}, {Filion}, {Gilbert}, {Escala}, {Mori}, {Kirihara}, {Tanaka}, {Ishigaki}, {Hayashi}, {Lee}, {Sharma}, {Kalirai}, \& {Lupton}}]{Ogami+2025}
{Ogami}, I., {Tanaka}, M., {Komiyama}, Y., {et~al.} 2025, \mnras, 536, 530

\bibitem[{{Preston} {et~al.}(2021){Preston}, {Collins}, {Rich}, {Ibata}, {Martin}, \& {Fardal}}]{Preston+2021}
{Preston}, J., {Collins}, M., {Rich}, R.~M., {et~al.} 2021, \mnras, 504, 3098

\bibitem[{{Sadoun} {et~al.}(2014){Sadoun}, {Mohayaee}, \& {Colin}}]{Sadoun+2014}
{Sadoun}, R., {Mohayaee}, R., \& {Colin}, J. 2014, \mnras, 442, 160

\bibitem[{{S{\'e}rsic}(1963)}]{Sersic1963}
{S{\'e}rsic}, J.~L. 1963, Boletin de la Asociacion Argentina de Astronomia La Plata Argentina, 6, 41

\bibitem[{{Takada} {et~al.}(2014){Takada}, {Ellis}, {Chiba}, {Greene}, {Aihara}, {Arimoto}, {Bundy}, {Cohen}, {Dor{\'e}}, {Graves}, {Gunn}, {Heckman}, {Hirata}, {Ho}, {Kneib}, {Le F{\`e}vre}, {Lin}, {More}, {Murayama}, {Nagao}, {Ouchi}, {Seiffert}, {Silverman}, {Sodr{\'e}}, {Spergel}, {Strauss}, {Sugai}, {Suto}, {Takami}, \& {Wyse}}]{Takada2014}
{Takada}, M., {Ellis}, R.~S., {Chiba}, M., {et~al.} 2014, \pasj, 66, R1

\bibitem[{{Tamura} {et~al.}(2024){Tamura}, {Yabe}, {Koshida}, {Moritani}, {Tanaka}, {Ishigaki}, {Ishizuka}, {Kamata}, {Allaoui}, {Arai}, {Arnouts}, {Barette}, {Barkhouser}, {Bergeron}, {Blanchard}, {Caplar}, {Carle}, {Chabaud}, {Chang}, {Chen}, {Chou}, {Cohen}, {Costa}, {Crauchet}, {de Almeida}, {de Oliveira}, {de Oliveira}, {Dohlen}, {dos Santos}, {Dobos}, {Ellis}, {Ertel}, {Fabricius}, {Ferreira}, {Furusawa}, {Gee}, {Garci{\'a}-Carpio}, {Gerasimov}, {Golebiowski}, {Gray}, {Gunn}, {Hahn}, {Hamano}, {Hammond}, {Harding}, {Hattori}, {Hayashi}, {He}, {Heckman}, {Hope}, {Hsu}, {Huang}, {Jaquet}, {Jeschke}, {Jespersen}, {Jing}, {Kackley}, {Karr}, {Kawanomoto}, {Kimura}, {Kirby}, {Koike}, {Komatsu}, {Koyama}, {Le Brun}, {Le Fur}, {Le Mignant}, {Lemson}, {Lin}, {Ling}, {Loomis}, {Lupton}, {Madec}, {Marchesini}, {Marrara}, {Medvedev}, {Mineo}, {Mitschang}, {Miyazaki}, {Morihana}, {Morishima}, {Murayama}, {Murray}, {Okamoto}, {Okita}, {Onodera}, {Passegger}, {Peebles}, {Price}, {Pyo}, {Ramos}, {Reiley}, {Reinecke},
  {Roberts}, {Rosa}, {Rousselle}, {Rubio}, {Schubert}, {Seiffert}, {Siegel}, {Smee}, {Sodr{\'e}}, {Strauss}, {Sunayama}, {Surace}, {Takada}, {Takagi}, {Tanaka}, {Tanaka}, {Thakar}, {Vibert}, {Wang}, {Wen}, {Werner}, {Wung}, {Yan}, {Yasuda}, \& {Yoshida}}]{Tamura2024}
{Tamura}, N., {Yabe}, K., {Koshida}, S., {et~al.} 2024, in Society of Photo-Optical Instrumentation Engineers (SPIE) Conference Series, Vol. 13096, Ground-based and Airborne Instrumentation for Astronomy X, ed. J.~J. {Bryant}, K.~{Motohara}, \& J.~R.~D. {Vernet}, 1309605

\end{thebibliography}

\end{document}